\documentclass[lettersize,journal]{IEEEtran}
\usepackage{amsmath, bm, mathabx}
\usepackage[colorlinks,citecolor=red,urlcolor=blue,bookmarks=false,hypertexnames=true]{hyperref}
\usepackage{color}
\usepackage{braket}
\usepackage{cite}
\usepackage{textgreek}
\usepackage{caption} 
\usepackage{orcidlink}

\begin{document}

\title{\LARGE\bf Development of KI-TWPAs for the DARTWARS project}

\author{F.~Ahrens \orcidlink{0009-0003-6609-795X},
        E.~Ferri \orcidlink{0000-0003-1425-3669},
        G.~Avallone \orcidlink{0000-0001-5482-0299},
        C.~Barone \orcidlink{0000-0002-6556-7556},
        M.~Borghesi \orcidlink{0000-0001-5854-8894},
        L.~Callegaro \orcidlink{0000-0001-5997-9960},
        G.~Carapella \orcidlink{0000-0002-0095-1434},
        A.~P.~Caricato \orcidlink{0000-0003-3512-1665},
        I.~Carusotto \orcidlink{0000-0002-9838-0149},
        A.~Cian \orcidlink{0000-0003-3497-7066},
        A.~D'Elia \orcidlink{0000-0002-6856-7703},
        D.~Di Gioacchino \orcidlink{0000-0002-1288-4742},
        E.~Enrico \orcidlink{0000-0002-2125-5200},
        P.~Falferi \orcidlink{0000-0002-1929-4710},
        L.~Fasolo \orcidlink{0000-0002-7537-7772},
        M.~Faverzani \orcidlink{0000-0001-8119-2953},
        G.~Filatrella \orcidlink{0000-0003-3546-8618},
        C.~Gatti \orcidlink{0000-0003-3676-1787},
        A.~Giachero \orcidlink{0000-0003-0493-695X},
        D.~Giubertoni \orcidlink{0000-0001-8197-8729},
        V.~Granata \orcidlink{0000-0003-2246-6963},
        C.~Guarcello \orcidlink{0000-0002-3683-2509},
        D.~Labranca \orcidlink{0000-0002-5351-0034},
        A.~Leo \orcidlink{0000-0002-0729-8386},
        C.~Ligi \orcidlink{0000-0001-7943-7704},
        G.~Maccarrone \orcidlink{0000-0002-7234-9522},
        F.~Mantegazzini \orcidlink{0000-0002-5620-2897},
        B.~Margesin \orcidlink{0000-0002-1120-3968},
        G.~Maruccio \orcidlink{0000-0001-7669-0253},
        R.~Mezzena \orcidlink{0000-0001-9891-0472},
        A.~G.~Monteduro \orcidlink{0000-0002-9265-134X},
        R.~Moretti \orcidlink{0000-0002-5201-5920},
        A.~Nucciotti \orcidlink{0000-0002-8458-1556}, 
        L.~Oberto \orcidlink{0000-0003-4073-0561},
        L.~Origo \orcidlink{0000-0002-6342-1430},
        S.~Pagano \orcidlink{0000-0001-6894-791X},
        A.~S.~Piedjou \orcidlink{0000-0003-2430-0086},
        L.~Piersanti \orcidlink{0000-0003-3186-3514},
        A.~Rettaroli \orcidlink{0000-0001-6080-8843},
        S.~Rizzato \orcidlink{0000-0002-3908-4796},
        S.~Tocci \orcidlink{0000-0002-5800-5408},
        A.~Vinante \orcidlink{0000-0002-9385-2127},
        M.~Zannoni \orcidlink{0000-0002-4495-571X}

\thanks{F.~Ahrens, A.~Cian, D.~Giubertoni, F.~Mantegazzini, and B.~Margesin are with
        Fondazione Bruno Kessler (FBK), I-38123 Trento, Italy, also with
        INFN - Trento Institute for Fundamental Physics and Applications, I-38123, Trento, Italy.
        (Corresponding author: F.~Ahrens, email: fahrens@fbk.eu)
        }
\thanks{E.~Ferri is with
        INFN - Milano Bicocca, I-20126, Milano, Italy.
        (Corresponding author: E.~Ferri, email: elena.ferri@mib.infn.it)
        }
\thanks{G.~Avallone, C.~Barone,  G.~Carapella, V.~Granata, C.~Guarcello, and S.~Pagano are with
        University of Salerno, Department of Physics, I-84084 Salerno, Italy and also with
        INFN - Napoli, Salerno group, I-84084 Salerno, Italy.
        }
\thanks{M.~Borghesi, M.~Faverzani, A.~Giachero, D.~Labranca, R.~Moretti, N.~Nucciotti, L.~Origo, and M.~Zannoni are with 
        University of Milano-Bicocca, Department of Physics, I-20126, Milano, Italy, also with
        INFN - Milano Bicocca, I-20126, Milano, Italy, and also with
        Bicocca Quantum Technologies (BiQuTe) Centre, I-20126 Milano, Italy. 
        }
\thanks{L.~Callegaro, and L.~Fasolo are with
        Istituto Nazionale di Ricerca Metrologica (INRiM), I-10135 Torino, Italy.
        }
\thanks{A.~P.~Caricato, A.~Leo, G.~Maruccio, A.~G.~Monteduro, and S.~Rizzato are with
        University of Salento, Department of Physics, I-73100 Lecce, Italy, and also with
        INFN - Lecce, I-73100 Lecce, Italy.
        }
\thanks{I.~Carusotto is with 
        Istituto Nazionale di Ottica - CNR I-38123 Trento, Italy, and also with
        University of Trento, Department of Physics, I-38123 Trento, Italy.
        }
\thanks{A.~D'Elia, D.~Di Gioacchino, C.~Gatti, C.~Ligi, G.~Maccarrone, A.~S.~Piedjou, L.~Piersanti, A.~Rettaroli, and S.~Tocci, are with
        INFN - Laboratori Nazionali di Frascati, I-00044, Frascati, Rome, Italy.
        }
\thanks{E.~Enrico, and L.~Oberto are with
        Istituto Nazionale di Ricerca Metrologica (INRiM), I-10135 Torino, Italy, and also with
        INFN - Trento Institute for Fundamental Physics and Applications, I-38123, Trento, Italy.
        }
\thanks{P.~Falferi, and A.~Vinante are with
        Istituto di Fotonica e Nanotecnologie - CNR, I-38123 Trento, Italy, also with
        Fondazione Bruno Kessler (FBK), I-38123 Trento, Italy, and also with
        INFN - Trento Institute for Fundamental Physics and Applications, I-38123, Trento, Italy.
        }
\thanks{G.~Filatrella is with
        University of Sannio, Department of Science and Technology, I-82100 Benevento, Italy and also with
        INFN - Napoli, Salerno group, I-84084 Salerno, Italy.
        }
\thanks{R.~Mezzena is with 
        University of Trento, Department of Physics, I-38123 Trento, Italy, and also with
        INFN - Trento Institute for Fundamental Physics and Applications, I-38123, Trento, Italy.
        }
}

\maketitle

This work is licensed to IEEE under the Creative Commons Attribution 4.0 (CCBY 4.0).\\
© 2024 IEEE. Personal use of this material is permitted. Permission from IEEE must be obtained for all other uses, in any current or future media, including reprinting/republishing this material for advertising or promotional purposes, creating new collective works, for resale or redistribution to servers or lists, or reuse of any copyrighted component of this work in other works

\begin{abstract}
Noise at the quantum limit over a broad bandwidth is a fundamental requirement for future cryogenic experiments for neutrino mass measurements, dark matter searches and Cosmic Microwave Background (CMB) measurements as well as for fast high-fidelity read-out of superconducting qubits. In the last years, Josephson Parametric Amplifiers (JPA) have demonstrated noise levels close to the quantum limit, but due to their narrow bandwidth, only few detectors or qubits per line can be read out in parallel. An alternative and innovative solution is based on superconducting parametric amplification exploiting the travelling-wave concept. Within the DARTWARS (Detector Array Readout with Travelling Wave AmplifieRS) project, we develop Kinetic Inductance Travelling-Wave Parametric Amplifiers (KI-TWPAs) for low temperature detectors and qubit read-out. KI-TWPAs are typically operated in a three-wave mixing (3WM) mode and are characterised by a high gain, a high saturation power, a large amplification bandwidth and nearly quantum limited noise performance. The goal of the DARTWARS project is to optimise the KI-TWPA design, explore new materials, and investigate alternative fabrication processes in order to enhance the overall performance of the amplifier. In this contribution we present the advancements made by the DARTWARS collaboration to produce a working prototype of a KI-TWPA, from the fabrication to the characterisation.

\end{abstract}


\section{The DARTWARS project}\label{sec:dtw}

\begin{figure}[!b]
    \centering
    \includegraphics[width = 0.45\textwidth]{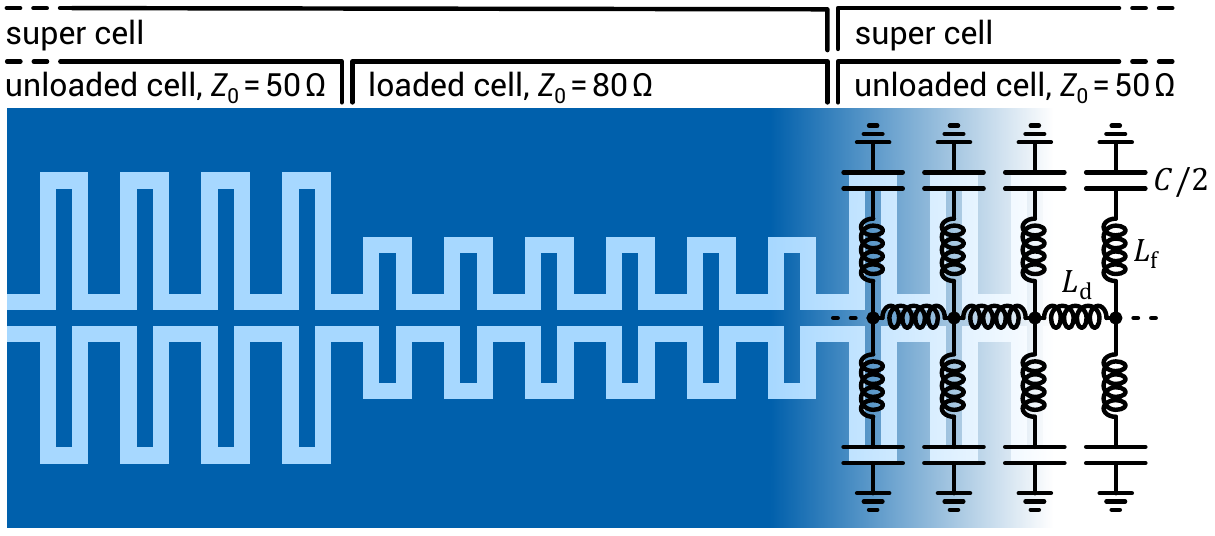}
    \caption{Schematics of the lumped element artificial transmission line. The dark (light) blue areas indicate the superconductor (substrate).}
    \label{fig:artificial_trans}
\end{figure}

The Detector Array Readout with Travelling Wave Amplifiers (DARTWARS) project~\cite{DARTWARSweb,Giachero2022} aims at developing a large bandwidth Kinetic Inductance Travelling-Wave Parametric Amplifier (KI-TWPA) with high gain and high saturation power, characterised by nearly quantum limited noise, suitable for low temperature detectors and qubit read-out \cite{Borghesi:2022uem}. A TWPA consists of a weakly dispersive transmission line with a phase-matched bandwidth, controlled by dispersion engineering, required to create exponential amplification and to prevent generation of harmonics~\cite{Eom2012}. The DARTWARS KI-TWPA design is inspired by the one proposed by M.~Malnou \cite{Malnou2021}. Each individual cell is composed of a coplanar waveguide (CPW), which consists of an inductive line with inductance $L$ and two interdigitated capacitors (IDC) stubs placed on either side of the central line. The stubs' length $l$ is adjusted to achieve a capacitance value $C$ such that $Z_0 = \sqrt{L/C} = 50\,\Omega$ (\textit{unloaded-cells}), as described in references~\cite{Chaudhuri2017, Malnou2021}. To achieve exponential amplification in the $4-8\,$GHz range, each \textit{super-cell} consists of $N_\mathrm{u}=60$ \textit{unloaded-cells} and $N_\mathrm{l}=6$ \textit{loaded-cells} (Fig.~\ref{fig:artificial_trans}). The length of the IDC stubs in the \textit{loaded cells} is adjusted to achieve a characteristic impedance of $Z_0 = \sqrt{L/C} = 80\,\Omega$. The gap and stub widths of the artificial transmission line are $1\,$\textmu m, while the stubs of the ground plane are $2\,$\textmu m  wide. 

\begin{figure}[t!]
    \centering
    \includegraphics[width = 0.45\textwidth]{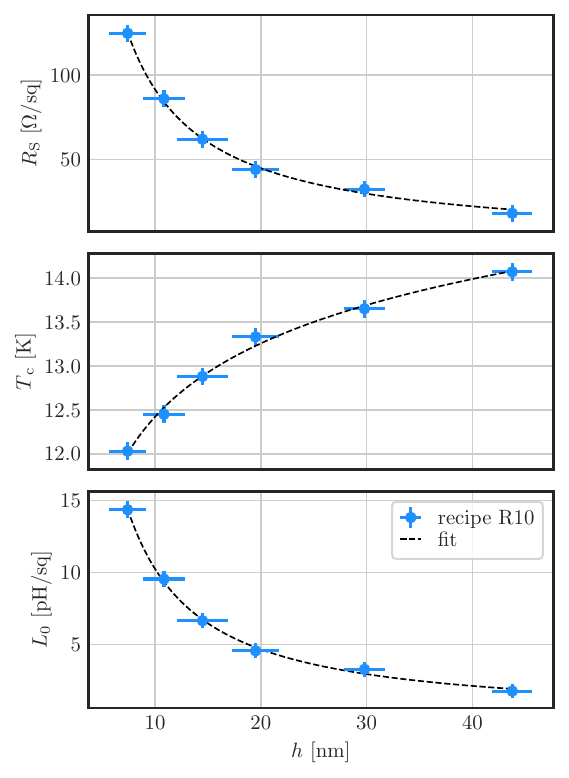}
    \caption{The sheet resistance $R_\mathrm{S}$, the critical temperature $T_\mathrm{c}$ and the expected kinetic inductance $L_0$ as a function of the film thickness $h$. The fit models are described in the text.}
    \label{fig:Tc_Rs_L0_vs_h}
\end{figure}

Each \textit{super-cell} is $330\,$\textmu m long, which is equivalent to half the wavelength ($\lambda/2$) of the pump signal in the frequency range of $f_\mathrm{p}=(8-12)\,$GHz. The  band gap is engineered to be just below the pump frequency $f_\mathrm{p}$: in the three-wave-mixing (3WM) process~\cite{Visser2016} this results in an amplification bandwidth centred around $f_\mathrm{p}/2$ ($4-6\,$GHz range in our case).
The first developed design consists of KI-TWPA prototypes patterned as meander-shaped lines composed of 523 super-cells (34,518 elementary cells) with a total length of around $17\,$cm. The expected gain is around $7-10\,$dB. While these gain levels may not be competitive, the results from these preliminary {\it half-size} amplifiers are crucial for refining the design in the final version. The {\it full-size} amplifier designs will be composed of 1,000 super-cells (66,000 elementary cells) patterned as double-arm spirals with a length of around $33\,$cm. This version will provide a $20\,$dB gain over a $2-3\,$GHz bandwidth centred at around $6\,$GHz. In this study, we present the fabrication processes developed for producing these devices and the first results from their characterisations.

\section{Fabrication of the prototype device}\label{sec:fab}

\begin{figure}[!b]
    \centering
    \includegraphics[width = 0.45\textwidth]{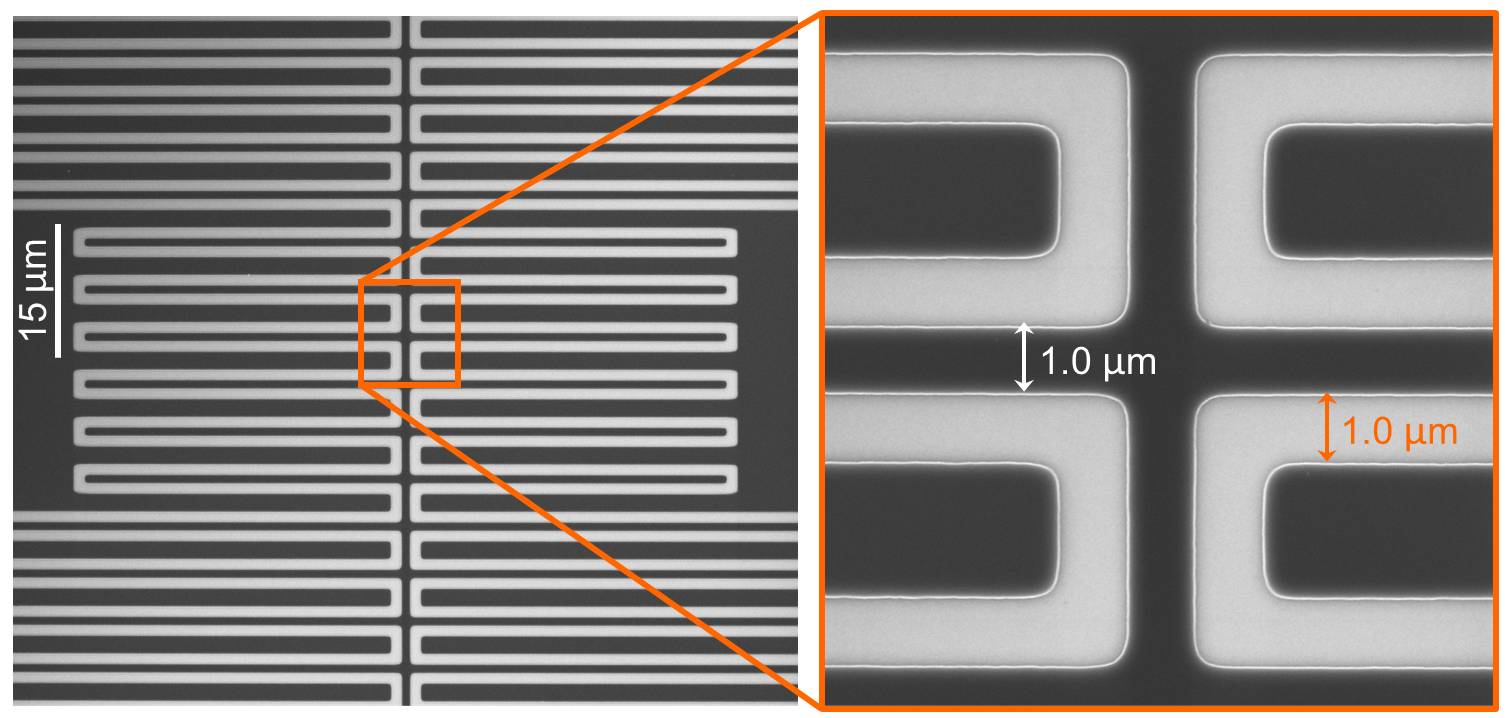}
    \caption{Post-microfabrication SEM images of the patterned artificial transmission line. Dark (light) areas correspond to the patterned NbTiN film (substrate).}
    \label{fig:SEM}
\end{figure}

The KI-TWPA prototype is made from a NbTiN thin film on a high-resistivity 6'' silicon wafer\footnote{FZ silicon wafers, diameter: 6", thickness: $625 \pm 15\,$\textmu m, dopant type: p, dopant: Boron, orientation: $<$100$>$ resistivity: $> 8000 \, \Omega / \mathrm{cm}$}. 
The film is deposited by means of magnetron sputtering\footnote{Kenosistec Physical Vapour Deposition, model KS 800 C Cluster} employing a Nb$_{80\%}$Ti$_{20\%}$ sputter target and adding nitrogen gas into the deposition chamber during the deposition process.

For the optimisation of the deposition we have prepared ten NbTiN thin film samples varying the gas pressure in the deposition chamber between $2\times10^{-3}\,$mbar and $3\times10^{-3}\,$mbar, the nitrogen flow into the deposition chamber between 4$\,$sccm\footnote{sccm = standard cubic centimeter per minute} and 8$\,$sccm, the chuck temperature during deposition between $300\,$°C and $400\,$°C, and the sputter power between $600\,$W and $1200\,$W, keeping the deposition time fixed at 6 minutes and the argon flow into the deposition chamber at 50$\,$sccm. After deposition, for each sample a cryogenic characterisation has been performed, determining the  the sheet resistance $R_\mathrm{S}$ and the critical temperature $T_\mathrm{c}$. Subsequently, the expected kinetic inductance $L_0$ of the NbTiN film can be inferred via the relation $L_0 = \frac{\hbar \cdot R_\mathrm{S}}{\pi \cdot T_\mathrm{c} \cdot k_\mathrm{B} \cdot 1.762}$ \cite{Bardeen1957}.

\begin{figure*}[!!t]
    \centering
    \includegraphics[width=0.65\textwidth]{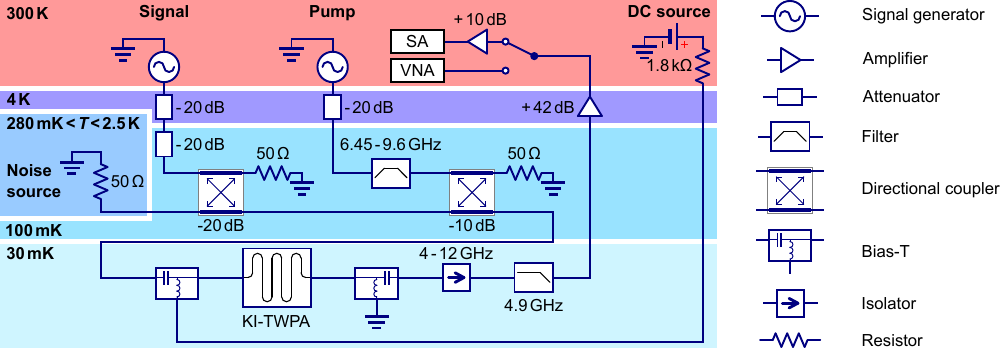}
    \caption{Schematics of measurement set-up for the determination of the KI-TWPA noise and gain performance.}
    \label{fig:setup}
\end{figure*}

The deposition recipe R10 with a sputter power of 600$\,$W, a pressure of $3 \times 10^{-3} \,$mbar, a nitrogen flow of 7$\,$sccm and a chuck temperature of 400$\,$\textdegree C yields both a sufficiently high sheet resistance of about 28$\, \Omega/$sq and a sufficiently high critical temperature of about 13.8$\,$K, leading to an expected kinetic inductance of about 2.1$\,$pH/sq for a deposition time of six minutes. Adjusting the film thickness by reducing the deposition time, it is possible to increase the sheet resistance and, consequently, the kinetic inductance of the film.
In order to exploit the film thickness as a fine tuning parameter to reach the desired kinetic inductance value, a high control on the film thickness, and thus, a precisely calibrated deposition rate are essential.
For this calibration, we have deposited six NbTiN films on silicon wafers featuring a silicon oxide layer, employing the deposition recipe R10, varying the deposition time. For the accurate determination of the film thickness, firstly, we have measured the wafers' pre-deposition oxide thickness employing optical interferometry, secondly, we have deposited the NbTiN films and patterned them by means of optical lithography and reactive ion etching, and thirdly, we have measured the etched step height via atomic force microscopy and the oxide thickness in the etched areas via optical interferometry. From the change in oxide thickness, we can infer the amount of over-etching and consequently, we deduce the film thickness from the measured step height and the measured over-etching. Fitting the film thickness as a function of deposition time, we have found a deposition rate of 5.2$\,$nm per minute for deposition recipe R10.

The samples also underwent a cryogenic characterisation that allowed to correlate the film properties with the film thickness. Fig.~\ref{fig:Tc_Rs_L0_vs_h} shows the resulting sheet resistance, critical temperature and expected kinetic inductance as a function of the film thickness.
 The dependence of the sheet resistance $R_\mathrm{S}$ on the film thickness $h$ matches the expectations based on Fuchs' model \cite{Fuchs1938}, as shown in the fit in Fig.~\ref{fig:Tc_Rs_L0_vs_h}, whereas the critical temperature $T_\mathrm{c}$ and the expected kinetic inductance $L_0$ as a function of the film thickness $h$ can be fit with phenomenological models $T_\mathrm{c}=\frac{AR_\mathrm{S}^{-B}}{h}$ \cite{Ivry2014} and $L_0(h) \propto h^{-\alpha}$ \cite{Zmuidzinas2012}, where $A$, $B$ and $\alpha$ are free parameters in the fits.
The sample with a film thickness of $h=10.8\,$nm features a sheet resistance of $R_\mathrm{S}=86.0\,\Omega$/sq and a critical temperature of $T_\mathrm{c}=12.5\,$K, leading to an expected kinetic inductance of about 9.5$\,$pH/sq.
We have measured the actual kinetic inductance employing a planar lumped-element microwave resonator made from a NbTiN film.
The capacitive element of the resonator is formed by an interdigital capacitance, whereas the inductive element is a narrow stripe of about 115 squares with negligible geometric inductance.
Comparing the measured resonance frequency to electromagnetic simulations, we found a kinetic inductance of $L_0=8.5\,$pH/sq.
The deviation of 11\,\% with respect to the expected value is within the systematic uncertainties arising from both the assumption of a BCS-like superconducting gap and the accuracy of the simulations.

Upon completion of the NbTiN thin film optimisation, we have fabricated the first KI-TWPA prototype device by means of NbTiN sputter deposition (recipe R10), stepper lithography and reactive ion etching. Fig.~\ref{fig:SEM} depicts scanning electron microscope (SEM) images of a section of the final device's artificial transmission line.

\section{Characterisation of the prototype device}\label{sec:char}

\begin{figure}[!b]
    \centering
    \includegraphics[width = 0.45\textwidth]{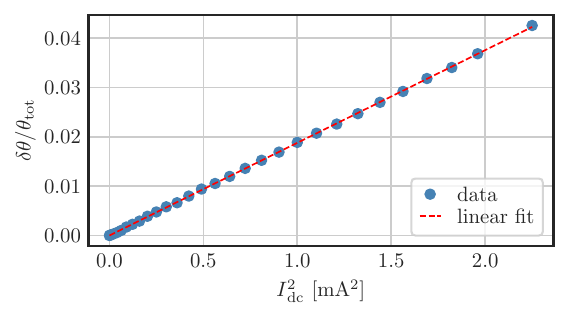}
    \caption{Non-linear phase shift as function of the dc current.}
    \label{fig:nonlinearity}
\end{figure}

The KI-TWPA prototypes ({\it half-size} amplifiers) firstly underwent an optical inspection and resistance measurements at room temperature. Subsequently, the devices have been characterised in liquid helium in order to measure the critical current $I_\mathrm{c}$, the band gap and the scaling current $I_*$. Devices with sufficiently high critical currents ($I_\mathrm{c}>1$\,mA) and pronounced band gap in the in transmission parameter $S_{21}$ have been tested at $20\,$mK in a dilution refrigerator assessing the devices' gain and noise performance. 

From the characterisation of the first device at a temperature of $4\,$K, the band gap and the non-linear phase shift as function of the dc current have been measured. The band gap is found to be at $7-7.7\,$GHz. The impedance of the artificial line is determined to be close to $50\,\Omega$, as estimated from the small Voltage Standing Wave Ratio (VSWR) below $6\,$GHz. 
Fig.~\ref{fig:nonlinearity} shows direct measurements of the non-linearity of the device. For this, a dc current $I_\mathrm{dc}$ is superimposed to a weak VNA rf signal and the output phase shift of the rf probe is measured as function of $I_\mathrm{dc}$. The signal frequency is set slightly above the band gap. The total phase shift along the device is estimated as $\theta_0 \approx N_\mathrm{sc}\, \pi$ where $N_\mathrm{sc}=523$ is the total number of \textit{super-cells}, with a small correction that can be applied to take into account the actual probe frequency with respect to the band gap center. A scaling current of $I_*=(5.3 \pm 0.1)\,$mA is extrapolated from the linear fit of the theoretical relative phase shift $\delta \theta/\theta_0$ as a function of $I_\mathrm{dc}^2$ \cite{Mantegazzini}. This value is in reasonable agreement with the theoretical value predicted by $3/2\,\sqrt{3}\,I_\mathrm{c}$ \cite{Zmuidzinas2012}, where $I_\mathrm{c} = (1.5 \pm 0.2)\,$mA is the critical current.

\begin{figure}[t]
    \centering
    \includegraphics[width = 0.45\textwidth]{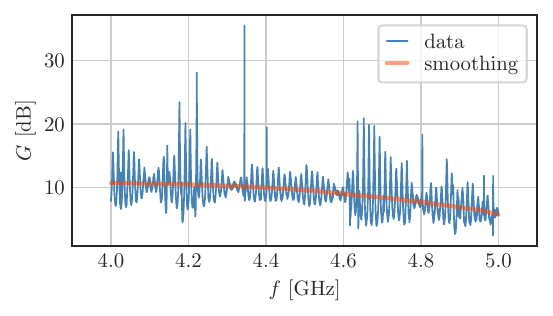}
    \caption{KI-TWPA gain at a dc current of $1.0\,$mA, a pump frequency of $8.05161\,$GHz and a pump power $-22.2\,$dBm.}
    \label{fig:gain}
\end{figure}

\begin{figure}[t]
    \centering
    \includegraphics[width = 0.45\textwidth]{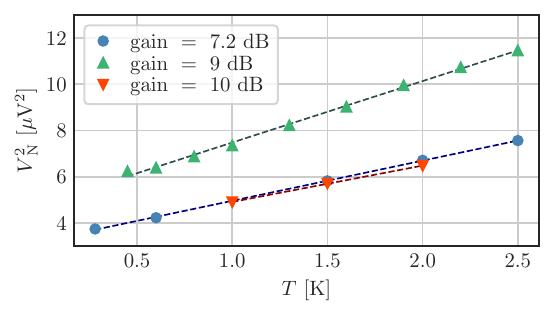}
    \caption{System noise measurements as a function of the noise source temperature.}
    \label{fig:noise}
\end{figure}

A second device, with suitable characteristics at room temperature and at $4\,$K, has been tested at $T=20\,$mK with the aim of investigating the gain and the noise performances of the KI-TWPA. The measurement set-up is depicted in Fig.~\ref{fig:setup}. The band gap frequency of this device was slightly higher compared to the one discussed above, i.e.~in the range  $7.5-8\,$GHz. The frequency range for measuring the gain and noise was selected from $4\,$GHz to $5\,$GHz, taking into account the constraints due to the high-pass cutoff of the isolator just below $4\,$GHz and the presence of a low-pass filter in the circuit, which is used to block the pump signal before reaching the cryogenic HEMT amplifier and prevent its saturation. Before the measurement, the dc current, the pump frequency, and the pump power were adjusted to optimise the gain while minimising the pump power in order to mitigate any potential heating issues. The KI-TWPA gain has been measured at different dc currents, pump frequencies and pump powers and it has been evaluated as the ratio of the transmitted signal with the pump switched on and off. An example of the KI-TWPA gain is reported in Fig. \ref{fig:gain}. The highest achieved gain was approximately $10\,$dB (averaged value). Even if this value is not competitive compared to the gain of $16.5\,$dB obtained by M.~Malnou \cite{Malnou2021}, we expect a gain of approximately $20\,$dB in our final full-size KI-TWPA layout, as the gain is expected to increase exponentially with the length of the transmission line. The ripples of the gain profile are caused by impedance mismatches within the device, and will be reduced in the full-size KI-TWPA due to an improved transmission line design. The measurement set-up depicted in Fig.~\ref{fig:setup} was carefully designed for the measurement of the KI-TWPA's noise performance, on the one hand meeting the input rf power constraints and, on the other hand, minimising the thermal noise coming from rf circuit components placed at temperatures higher than $20\,$mK. The noise source consists of a 50$\,\Omega$ resistance anchored to the cold plate at a temperature close to $100\,$mK and is set to different temperatures in the range from $0.28\,$K to $2.5\,$K. The output signal of the amplifier is measured with a spectrum analyser. The noise measurements are performed at three different KI-TWPA gains as shown in Fig.~\ref{fig:noise}. The experimental points are fit with $V_\mathrm{n}^2 = A+ BT$ and the system noise temperature $T_\mathrm{sys}$ is evaluated as the ratio between the intercept and the slope. The bare noise contribution of the KI-TWPA can be extracted considering the following equation:

\begin{equation}\label{eq:noise_T}
    N_\mathrm{out} =G_\mathrm{H}\,T_\mathrm{H} + (G_\mathrm{K}\,F_2\,G_\mathrm{H})\,T_\mathrm{n} + F_1\,F_2\,G_\mathrm{H} (G_\mathrm{K}+ G_\mathrm{K} -1)\,T .
\end{equation}

Here, $G_\mathrm{K}$, $G_\mathrm{H}$, $T_\mathrm{n}$, and $T_\mathrm{H}$ represent the gain and noise temperature of the KI-TWPA and cryogenic HEMT amplifier, respectively. $F_1$ denotes the attenuation of the transmission lines and other microwave devices between the resistance and the KI-TWPA, while $F_2$ represents the attenuation between the KI-TWPA and the HEMT. Note that the contribution $G_\mathrm{K} -1$ in the last term arises from the idler. The experimental results for various KI-TWPA gain settings are reported in table \ref{TAB:Noise_Temp}, the best value being 2.5 photons ($500\,$mK at $4\,$GHz), which is close to the quantum limit of 1/2 photon ($100\,$mK at $4\,$GHz). This result is obtained with a gain of approximately $7.2\,$dB, which is not the maximum gain achieved. One possible explanation is that higher gains require increased pump power, leading to heating and a higher density of quasi-particles in the device.

\begin{table}
\centering
  \begin{tabular}{ |p{1.5cm}||p{1.0cm}|p{1.0cm}|p{1.0cm}| }
 \hline
 \multicolumn{4}{|c|}{Noise measurements results} \\
 \hline
 $G_\mathrm{K}$ (dB)   & 7.2  & 9.0  & 10   \\
 $T_\mathrm{sys}$ (K) & 1.78 & 1.72 & 2.06 \\
 $T_\mathrm{n}$ (K)    & 0.5  &0.7   & 1.0  \\
 $N_\mathrm{q}$   & 2.5   & 3.4   & 4.9   \\
 \hline
\end{tabular}  
\caption{Noise temperature and equivalent noise quanta $N_\mathrm{q}$ as a function of the KI-TWPA gain.}
\label{TAB:Noise_Temp}
\end{table}

\section{Conclusions}\label{sec:conclusion}
We have optimised the sputter deposition of NbTiN thin films and the microfabrication process in order to fabricate KI-TWPA prototype devices consisting of artificial transmission lines. Having gained precise control on the deposition parameters, we are able to exploit the film thickness as fine tuning parameter for the kinetic inductance value of the film. The first KI-TWPA prototypes have been fabricated and characterised, showing a gain of approximately $10\,$dB, aligned with the expected values. Furthermore, they achieve a noise level near to the standard quantum limit. In conclusion, even though the half-size KI-TWPA prototype does not yet match the state-of-the-art, the results obtained with this device are very promising. The microfabrication and design can be easily adapted to an optimised full-size device, which can potentially reach beyond state-of-the-art performance.

\section*{Acknowledgement}
This work is supported by DARTWARS (funded by INFN CSN5 and by EU’s H2020-MSCA Grant Agreement No. 101027746) and also by NQSTI through the PNRR MUR Project, Grant PE0000023-NQSTI. We acknowledge the support of the FBK cleanroom and the useful discussions with Jiansong Gao, Michael Vissers, and Jordan Wheeler. CB and SP acknowledge support from University of Salerno (projects FRB19PAGAN, FRB20BARON and FRB22PAGAN).

\bibliographystyle{IEEEtran}
\bibliography{main}

\end{document}